\documentclass[useAMS,usenatbib]{mn2e}

\usepackage{graphicx}
\usepackage{txfonts}
\usepackage{natbib}
\usepackage{epsfig}
\usepackage{enumerate}

\newcommand{\ems}{1FGL J1227.9$-$4852}
\newcommand{\fermi}{{\em Fermi}}
\newcommand{\xss}{XSS J12270$-$4859}
\newcommand{\atca}{J122806$-$485218}

\title[The bright unidentified $\gamma$-ray source 1FGL J1227.9$-$4852]{The bright unidentified $\gamma$-ray source 1FGL J1227.9$-$4852:\\ Can it be associated with an LMXB?}
\author[A. B. Hill]{A. B. Hill$^{1}$\thanks{E-mail:adam.hill@obs.ujf-grenoble.fr} 
, A. Szostek$^{1,3}$, S. Corbel$^{2}$, F. Camilo$^{4}$, R. H. D. Corbet$^{5,6}$, R. Dubois$^{7}$,  
\newauthor
G. Dubus$^{1}$, P. G. Edwards$^{7}$, E. C. Ferrara$^{5}$, M. Kerr$^{7}$, E. Koerding$^{2}$, D. Kozie{\l}$^{3}$, {\L}. Stawarz$^{9,3}$\\
$^{1}$Laboratoire d'Astrophysique de Grenoble, UMR 5571 Universit\'e Joseph Fourier Grenoble I / CNRS, BP 53, 38041 Grenoble, France\\
$^{2}$Universit\'e Paris 7 Denis Diderot and Service d'Astrophysique, UMR AIM,  CEA Saclay, F-91191 Gif sur Yvette, France\\
$^{3}$Astronomical Observatory, Jagiellonian University, Orla 171, 30-244 Krak\'ow, Poland\\
$^{4}$Columbia Astrophysics Laboratory, Columbia University, New York, NY 10027, USA\\ 
$^{5}$NASA Goddard Space Flight Center, Greenbelt, MD 20771, USA\\ 
$^{6}$Center for Space Science and Technology, University of Maryland Baltimore County, Baltimore, MD 21250, USA\\ 
$^{7}$W. W. Hansen Experimental Physics Laboratory, Kavli Institute for Particle Astrophysics and Cosmology, Department of Physics and SLAC National Accelerator\\ Laboratory, Stanford University, Stanford, CA 94305, USA\\ 
$^{8}$CSIRO Astronomy and Space Science, P.O. Box 76, Epping NSW 1710, Australia\\ 
$^{9}$Institute of Space and Astronautical Science, JAXA, 3-1-1 Yoshinodai, Chuo-ku, Sagamihara, Kanagawa 252-5210, Japan\\}

\begin{document}

\date{Accepted 2010 xxxx}

\pagerange{\pageref{firstpage}--\pageref{lastpage}} \pubyear{2002}

\maketitle

\label{firstpage}

\begin{abstract}
We present an analysis of high energy (HE; 0.1--300 GeV) $\gamma$-ray observations of \ems\ with the {\em Fermi} Gamma-ray Space Telescope, follow-up radio observations with the Australia Telescope Compact Array, Giant Metrewave Radio Telescope and Parkes radio telescopes of the same field and follow-up optical observations with the ESO VLT.  We also examine archival \emph{XMM-Newton} and \emph{INTEGRAL} X-ray observations of the region around this source.  The $\gamma$-ray spectrum of \ems\ is best fit with an exponentially cutoff power-law, reminiscent of the population of pulsars observed by \fermi.  A previously unknown, compact radio source within the 99.7\% error circle of \ems\ is discovered and has a morphology consistent either with an AGN core/jet structure or with two roughly symmetric lobes of a distant radio galaxy.  A single bright X-ray source \xss, a low-mass X-ray binary, also lies within the \ems\ error circle and we report the first detection of radio emission from this source.  The potential association of \ems\ with each of these counterparts is discussed.  Based upon the available data we find the association of the $\gamma$-ray source to the compact double radio source unlikely and suggest that \xss\ is a more likely counterpart to the new HE source.  We propose that \xss\ may be a millisecond binary pulsar and draw comparisons with PSR J1023+0038.
\end{abstract}


\section{Introduction}
The first year (1FGL) catalogue of the \emph{Fermi} Large Area Telescope (LAT) \citep{1FGLcat} lists in excess of 1400 $\gamma$-ray sources (0.1--300 GeV) across the entire sky of which more than 600 are not associated with a known $\gamma$-ray emitting object or source class and are hence considered unidentified.  In an attempt to identify interesting sources for follow-up, we cross-correlated the 1FGL catalogue with the latest hard X-ray catalogues from {\em Swift} and {\em INTEGRAL} \citep{2010A&A...510A..48C, 2010ApJS..186....1B}.  The cross-correlation yields 50--60 sources which are good candidates for having hard X-ray counterparts.  The vast majority of these are blazars \citep[see][]{1LAC} or bright well known high-energy Galactic sources such as the the Crab pulsar; this is in agreement with the early findings of \citet{2009ApJ...706L...7U} who correlated the 3-month \emph{Fermi} Bright Source List with hard X-ray catalogues.  

One unidentified \emph{Fermi} source associated through the cross-correlation, 1FGL J1227.9$-$4852, does not fall into these `common' source classes; it correlates with the hard X-ray source \xss\ \citep{2010A&A...510A..48C, 2010ApJS..186....1B}, an unusual low mass X-ray binary\footnote{\xss\ was originally classified as an intermediate polar cataclysmic variable \citep{2008A&A...487..271B}.  However, recent independent studies have cast doubt on this classification \citep[see][]{2009PASJ...61L..13S, 2010A&A...515A..25D}.}(LMXB) \citep{2009PASJ...61L..13S, 2010A&A...515A..25D}.  The 1FGL source location is approximately 14$^\circ$ above the Galactic plane at R.A.~=~12$^{\rm h}$27\fm9, Dec.~=~$-$48\degr52\farcm7 (J2000) with a 95\% error radius of $\sim$6.1\arcmin.  The 1FGL catalogue lists it as having no known high energy counterpart, however within the 95\% error circle are the previously mentioned hard X-ray and two 2MASS catalogue objects classified as regular galaxies. There is no known radio source within the error circle, however there is a single SUMSS source just outside of the 95\% radius, SUMSS J122820$-$485537. \citet{2010A&A...515A..25D} noted the proximity of SUMSS J122820$-$485537 to \ems. However, as shown in $\S$~\ref{sec:LAT} the refined LAT error circle completely discounts this SUMSS source as a potential counterpart.

To date, three high-mass X-ray binaries (HMXBs) have been detected by \fermi-LAT: LS~I~+61$^\circ$~303 \citep{2009ApJ...701L.123A}, LS 5039 \citep{2009ApJ...706L..56A} and Cyg X-3 \citep{2009Sci...326.1512F}.  All of these systems contain $>10$M$_{\odot}$ donor stars. The first two sources are seen as persistent, bright variable $\gamma$-ray sources, while the latter is a microquasar observed as a HE transient.  The location of \ems\ $\sim$14$^\circ$ off the Galactic plane rules out the possibility of it being a distant HMXB due to the absence of massive stars at high Galactic latitudes.

\xss\ makes for an intriguing potential counterpart and, if associated with the $\gamma$-ray source, would be the first LMXB detected by \emph{Fermi}.  This X-ray source was recently the subject of a detailed study by \citet{2010A&A...515A..25D} who performed an analysis of targeted {\em XMM-Newton} and optical V band observations combined with archival {\em INTEGRAL} and {\em RXTE} observations; they also noted its spatial coincidence with a 1FGL catalogue source.  While there are a number of X-ray sources in the field \xss\ is far and away the brightest and is characterised by \citet{2010A&A...515A..25D} as a persistent X-ray source which exhibits flaring and dipping on timescales of 10s of minutes, as originally reported by \citet{2009PASJ...61L..13S} from \emph{Suzaku} observations.  The spectrum is reported to be best represented by an absorbed power law with a hydrogen column density of $\sim$10$^{21}$cm$^{-2}$.  No evidence of a cutoff in the spectrum is found up to 55 keV.  The unabsorbed 0.2--100 keV flux is given as (4.2 $\pm$ 0.2) $\times$10$^{-11}$ erg cm$^{-2}$~s$^{-1}$ giving a luminosity of $\sim$5.0~$\times$~10$^{33}$~erg~s$^{-1}$ if the source is 1 kpc distant. \citet{2010A&A...515A..25D} also report optical V band observations made with the Rapid Eye Mount (REM) robotic telescope at the ESO, La Silla observatory.  They find a periodic signal which yields a period of 4.32$\pm$0.01 hours when fit with a sinusoid. They note that the weak quasi-sinusoidal variability observed in the {\em XMM-Newton} quiescent X-ray light curve is consistent with the 4.32 hr period and could be ascribed to a binary orbit.  They conclude that the source is likely a peculiar, low-luminosity low mass X-ray binary with a possible orbital period of 4.32 hr. \xss\ has also been reported as a hard X-ray source detected at energies up to 100 keV by {\em INTEGRAL} \citep{2010ApJS..186....1B} and {\em Suzaku} \citep{2009PASJ...61L..13S}.

We present here a detailed analysis of the {\em Fermi} HE $\gamma$-ray data and additional follow-up radio observations of the region.  Together with archival X-ray and hard X-ray data we attempt to explain the underlying nature of this mysterious source.

\section{Observations and data reduction}

\subsection{{\em Fermi}-LAT observations of 1FGL J1227.9$-$4852}\label{sec:LAT}

The LAT is the primary instrument onboard \fermi; it is an electron-positron pair production telescope, featuring
solid state silicon trackers and cesium iodide calorimeters, sensitive to photons from $\sim 20$ MeV to $>300$ GeV \citep{2009ApJ...697.1071A}.  The observatory operates principally in survey mode; the telescope is rocked north and south on alternate orbits to provide more uniform coverage so that every part of the sky is observed for $\sim$30 minutes every 3 hours.
 
The LAT analysis dataset spanned 2008-08-04 UTC 15:43:37 to 2010-09-15 UTC 04:46:08. The data were reduced and analyzed using the {\em Fermi} Science Tools
v9r15 package\footnote{See the FSSC website for details of the Science Tools:
  http://fermi.gsfc.nasa.gov/ssc/data/analysis/}. The standard onboard
filtering, event reconstruction, and classification were applied to
the data \citep{2009ApJ...697.1071A}, and for this analysis the
high-quality (``diffuse") photon event class is used.  Throughout the
analysis, the ``Pass 6 v3 Diffuse'' (P6\_V3\_DIFFUSE) instrument
response functions (IRFs) are applied. Time periods when the region around \ems\ was observed at a zenith angle greater than 105\degr and for observatory rocking angles of greater than 43\degr before MJD 55077 and greater than 52\degr after MJD 55077 were also excluded to avoid contamination from the Earth limb photons.

\begin{figure}
\centering
\includegraphics[width=0.95\linewidth, clip]{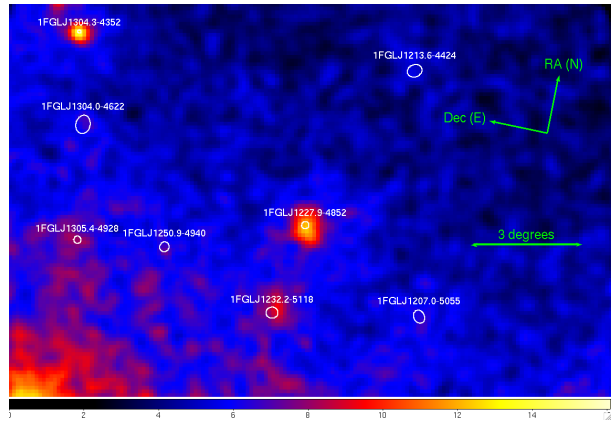}
\caption{The smoothed counts map for 100~MeV--300~GeV of the region around 1FGL J1227.9$-$4852 using the $\sim$25.5 month dataset described in the text; the white ellipses denote the 95\% error circles of the 1FGL sources within the field \citep{1FGLcat}.  An 0\fdg3 gaussian smoothing function was applied to the 0\fdg1 bins.}
\label{fig:FermiImage}
\end{figure}

\begin{figure}
\centering
\includegraphics[width=0.95\linewidth, clip]{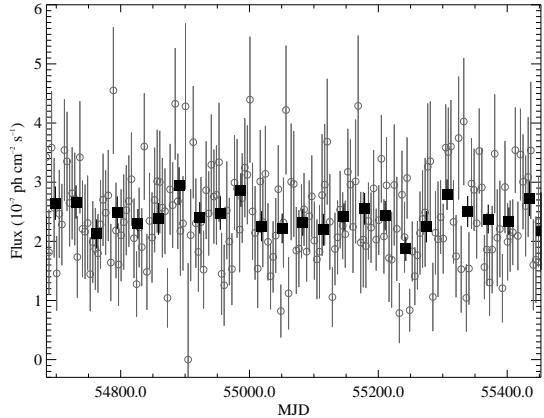}
\caption{The {\em Fermi}-LAT 0.1--300 GeV aperture photometry light curve of 1FGL J1227.9$-$4852 for the first $\sim$25.5 months of observations; no background was subtracted and a 1$^\circ$ aperture was used.  The grey circles indicate 4 day flux bins; the black squares indicate 32 day flux bins.}
\label{fig:lc}
\end{figure}

\subsubsection{Imaging and timing analysis}\label{timing}
The photon count map of the region surrounding \ems\ is shown in Figure~\ref{fig:FermiImage}.  The source can be seen to be bright and isolated.  The source is localised using the {\tt pointfit} likelihood tool \citep[see][]{1FGLcat}. For this and subsequent likelihood analyses a spectral-spatial model containing point and diffuse sources was created and the parameters obtained from a simultaneous maximum likelihood fit to the data.  All point sources within a 15\degr\ region around \ems\ were included in the model (36 sources listed in the 1FGL catalogue) as were the Galactic and isotropic diffuse contributions.  The point sources were modelled by a simple power-law function and the spectral parameters were frozen to those reported in the 1FGL catalogue except those sources within 3$^\circ$ of \ems\ for which the spectral parameters were simultaneously fit.  The models used for the Galactic diffuse emission ($gll\_iem\_v{\it 02}.fit$) and isotropic backgrounds ($isotropic\_iem\_v02.txt$) were those currently recommended by the LAT team\footnote{Descriptions of the models are available from the FSSC: http://fermi.gsfc.nasa.gov/}.  \ems\ is localised at R.A.~=~12$^{\rm h}$27\fm7, Dec.~=~$-$48\degr53\farcm0 (J2000) with a 95\% error radius of 2.9\arcmin; this is compatible with, but much more refined than, the reported position in the 1FGL catalogue \citep{1FGLcat}.  
\
A LAT aperture photometry light curve was extracted for the source using the {\tt gtbin} tool and the exposure correction was then calculated using the {\tt gtexposure} tool; it should be noted that these tools perform no background subtraction in the light curve generation. A 0.1--300 GeV light curve was produced with 4-day binning and using a 1$^\circ$ aperture. The light curve is shown in Figure~\ref{fig:lc} and shows the source to be very stable with no indications of variability on timescales longer than the 4-day binning; the light curve can be adequately modelled with a constant flux.

A light curve was constructed on a much shorter binning time of 1000s to search for the 4.32 hr periodicity reported by \citep{2010A&A...515A..25D} in their X-ray observations of \xss.  Using the Lomb-Scargle periodogram method \citep{1976Ap&SS..39..447L, 1982ApJ...263..835S} no significant periodicity was detected at the 4.32 hr period or any other period longer than 1000s.  
The LAT source \ems\ was one of the $\sim$650 sources included in the pulsar search described in \citet{2010ApJ...725..571S}. Using approximately one year of sky survey mode observations, a search was carried out using diffuse class events above 300 MeV, within a 1.5$^\circ$ radius of the 1FGL Catalog position. A time differencing window of 3 days was used and the standard parameter space (up to 64 Hz) was searched, as described in \citet{2009Sci...325..840A} and \citet{2010ApJ...725..571S}. No significant candidates were found in the search; although these searches would not be sensitive to a millisecond pulsar.

\subsubsection{Spectral fitting}

\begin{figure}
\centering
\includegraphics[width=0.95\linewidth, clip]{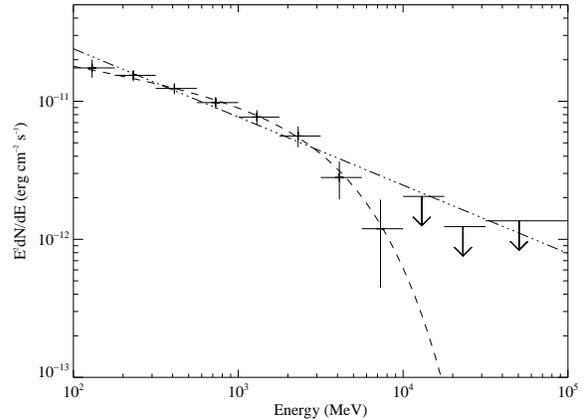}
\caption{The LAT SED of 1FGL J1227.9$-$4852.  \fermi\ data points are from
  likelihood fits in each energy bin and are indicated by the black data points; 95\% upper limits are calculated for the highest 3 energy bins indicated by arrows. The dashed line indicates the best fit exponential cut-off power law; the dot-dashed line indicates the simple power law fit, see the text for details.}
\label{fig:fermiSpec}
\end{figure}

The {\tt gtlike} likelihood fitting tool was used to perform the
spectral analysis using the model described in \S~\ref{timing}.
1FGL J1227.9$-$4852 was initially fit with a simple power-law model yielding a photon index of $\Gamma \sim 2.49$.  This is compatible with the spectral shape reported in the 1FGL catalogue \citep[$\Gamma = 2.45 \pm 0.07$,][]{1FGLcat}.  Subsequent investigation found that the source was best represented by an exponentially cutoff power-law model of the form $E^{-\Gamma}\exp{[-(E/E_{\rm cutoff})]}$.  The best fit spectral parameters are a photon index $\Gamma$ = 2.21 $\pm$ 0.09~(stat) $^{+0.35}_{-0.4}$~(syst) with a cutoff at $E_{\rm cutoff}$ = 4.1 $\pm$ 1.3~(stat) $^{+3.3}_{-1.4}$~(syst) GeV and with an integral flux of (8.8 $\pm$ 0.7~(stat) $^{+5.1}_{-3.5}$~(syst)) $\times$10$^{-8}$ ph cm$^{-2}$s$^{-1}$ (0.1--300 GeV).  This gives a luminosity of $\sim$4.9~$\left(\frac{d}{1 {\rm kpc}}\right)^{2}$~$\times$~10$^{33}$~erg~s$^{-1}$ for the source distance $d$.  The maximum likelihood exponential cutoff spectral model has a likelihood test statistic \citep{1996ApJ...461..396M} value of $\sim$820.8 equating to a formal detection significance of $>$27$\sigma$. Comparing the results of the maximum likelihood fit of the two spectral models discussed above finds that $-2\Delta\log(Likelihood) = 15.74$ which implies that the spectral model with the cutoff improves the maximum likelihood at the $\sim$4$\sigma$ level over the simple power law.  Figure~\ref{fig:fermiSpec} shows the LAT data; Table~\ref{tab:spec} presents the fit results of the power-law and exponentially cutoff power-law models.

The possibility of a broken power-law model representing the data was also investigated as this is a common model used to describe the high energy spectra of blazars where a break appears around a few GeV \citep[e.g.][]{2009ApJ...699..817A}. However, a reasonable fit was not achieved and no significant increase in TS was achieved using this spectral model.

A number of effects are expected to contribute to the systematic
errors quoted above. Primarily, these are uncertainties in the effective area and
energy response of the LAT as well as background contamination. The systematics are
currently estimated by using outlier IRFs that bracket the
nominal ones in effective area. The outlier IRFs are defined by envelopes above
and below the P6\_V3\_DIFFUSE IRFs by linearly connecting differences of (10\%,
5\%, 2\%) at log(E$/$MeV) of (2, 2.75, 4) respectively.

\begin{table}
\centering
\caption{The results of the spectral fits to the \ems\ using the {\em Fermi}-LAT 0.1--300 GeV data. Only statistical errors are quoted see text for a discussion of systematic error estimates.}
\begin{tabular}{|l|cc}
\hline
Parameter & Power law &  Cutoff \\
 &   & power law \\
\hline
\hline
Photon index  &  2.49$\pm$0.04 &  2.21 $\pm$ 0.09 \\
Cutoff energy (GeV)	&	- &  4.1 $\pm$ 1.3 \\
Flux $>$100 MeV & 10.1$\pm$0.7	&	 8.8 $\pm$ 0.7\\
10$^{-8}$ ph cm$^{-2}$s$^{-1}$ & & \\
\hline
\end{tabular}
\label{tab:spec}
\end{table}

\subsection{ATCA \& GMRT Radio observations}
In order to investigate the \fermi-LAT source region, we conducted three continuum radio observations with the Australia  Telescope Compact Array (ATCA) located in Narrabri, New South Wales, Australia. The ATCA synthesis  telescope is an east-west array consisting of six 22 m antennas. The ATCA uses orthogonal polarized  feeds and records full Stokes parameters. We carried out the observations with the upgraded Compact Array Broadband Backend (CABB) that provides a new broadband backend system for the ATCA and increases the maximum bandwidth from 128 MHz to 2 GHz.

The observations on 2009 October 2 and November 6 were performed at two frequency bands simultaneously, with central frequencies at 5.5 GHz and 9  GHz. On 2009 October 13, both frequencies were set to 2.45 GHz, using the maximum bandwidth available at the time of 500 MHz.  The ATCA was in a very compact configuration on 2009 October 2 (H75) and 2009 October 13 (H168) and in an intermediate configuration (1.5B) on 2009 November 6.  The observation log is given in Table~\ref{tab:ATCA}. 

The amplitude and band-pass calibrator was PKS~1934$-$638, and the antenna's gain and phase  calibration, as well as the polarization leakage, were derived from regular observations of the  nearby ($\sim$3.3 degrees away) calibrator PMN~1215$-$457.  The editing, calibration, Fourier transformation, deconvolution, and image analyses were performed using the MIRIAD software package  \citep{1995ASPC...77..433S,MiriadUG}. Cleaning was carried out using a combination of multi-frequency \citep{1994A&AS..108..585S} and standard clean algorithms.

Low frequency observations were performed with the Giant Metrewave Radio telescope (GMRT) near Pune, India. The GMRT consists of 30 fully steerable parabolic dishes of 45~m diameter each spread over distances of up to 25 km. The field was observed on 2010 May 10 at 640 MHz and 240 MHz and on 2010 May 13 at 1400 MHz; see Table~\ref{tab:ATCA}. The data were converted into FITS format and reduced using standard AIPS procedures.  The primary flux calibration was done using 3C286, which was also used as a bandpass calibrator. All observations were phase referenced to PKS 1151$-$34.

\begin{figure}
\centering
\includegraphics[width=0.95\linewidth,clip=]{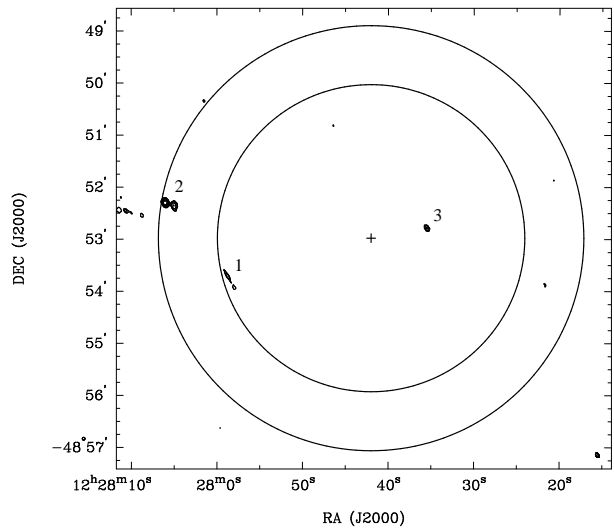}
\caption{Radio emission at 5.5 GHz at the location of \ems\ from the ATCA observation on 2009 November 6; the circles indicate the 95\% and 99.7\% \emph{Fermi}-LAT error circles. Contours are at $-3$,3,5,10,20,30 \& 50 times the level of 0.20 mJy/beam. The synthesized beam (in the lower right corner) is 5.6\arcsec $\times$ 3.5\arcsec, with the major axis at a position angle of 28$^\circ$.  Three distinct radio sources are found within the error box.}
\label{fig:atca-all}
\end{figure}

\begin{figure}
\centering
\includegraphics[height=0.85\linewidth,angle=0,clip=]{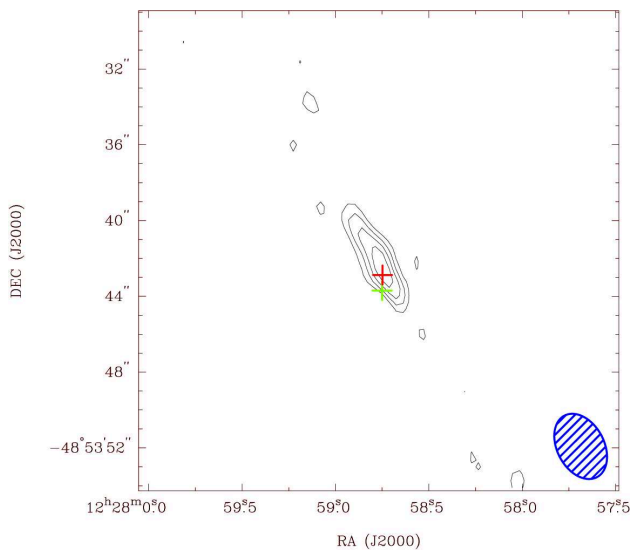}
\caption{Radio emission at 9 GHz around the location of \xss\ from the ATCA observation on 2009 November 6. Contours are at -3, 3, 4, 5, and 6 times the r.m.s. noise level of 0.20 mJy/beam. The synthesized beam (in the lower right corner) is 3.7\arcsec $\times$ 2.5\arcsec, with the major axis at a position angle of 28$^\circ$. The red cross marks the location of the optical counterpart of \citet{2006A&A...459...21M}; the green cross marks the \emph{XMM-Newton} X-ray location.}
\label{fig:atca-xss}
\end{figure}

\begin{figure*}
\centering
\includegraphics[width=0.85\linewidth,clip=]{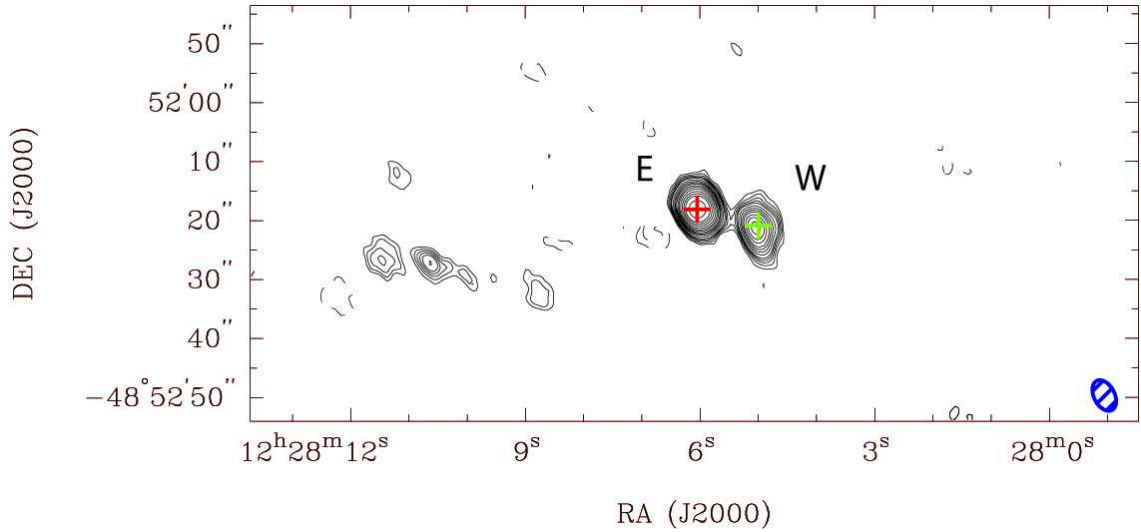} 
\caption{Radio emission at 5.5 GHz around the location of  J122806$-$485218 from the ATCA observation on 2009 November 6. Contours are at -3, 3, 4, 5, 6, 7, 9, 11, 13, 15, 18, 21, 25, 30, 35, 40 and 50 times the r.m.s. noise level of 0.15 mJy/beam. The synthesized beam (in the lower right corner) is 5.6\arcsec $\times$ 3.5\arcsec, with the major axis at a position angle of 28$^\circ$. The red cross marks the location of the radio core.}
\label{fig:radio2}
\end{figure*}

Within the 99.7\% error circle of \ems\ three distinct radio sources are detected in the ATCA observations as shown in Figure~\ref{fig:atca-all}.  All three of these radio sources are new discoveries and are not reported in any current radio catalogues; the subsequently reported source locations are all based upon the data from observation 3.

\textbf{Source 1}: A faint radio source is detected above the 5$\sigma$ level at a location compatible with the low-mass X-ray binary \xss\ \citep{2006A&A...459...21M} and is shown in Figure~\ref{fig:atca-xss}.  The radio source is located at R.A.~=~12$^{\rm h}$27$^{\rm m}$58.$^{\!\rm s}$8, Decl.~=~$-$48\degr53\arcmin42\farcs1 (J2000) with an error of 0\farcs5.  The source flux density is 0.18~$\pm$~0.03 mJy at 5.5 GHz and 0.14~$\pm$~0.03 mJy at 9 GHz with no detection in any of the GMRT bands. It has a spectrum F$_\nu\sim\nu^{\alpha}$ with $\alpha=-0.5~\pm$~0.6.  This implies an optically thin spectrum but could be consistent with a flat spectrum.  This is the only radio source within the error circle of \ems\ that has an obvious X-ray counterpart.

\textbf{Source 2}: The brightest radio source in the field, hereafter J122806$-$485218 (named after the brighter of the 2 components), is a double source comprised of a brighter eastern component, E, (possibly a radio core) and a weaker western component, W (most likely a jet or a one-sided lobe), as illustrated in Figure~\ref{fig:radio2}.  E is located at R.A.~=~12$^{\rm h}$28$^{\rm m}$06.$^{\!\rm s}$04 ($\pm$0.$\!^{\rm s}$09), Decl.~=~$-$48\degr52\arcmin18\farcs05 ($\pm$0\farcs10) (J2000).  The observed flux densities in the 5.5 and 9 GHz bands are 1.36~$\pm$~0.03 mJy and 1.03~$\pm$~0.03 mJy respectively; E is also detected in the GMRT 1400 MHz band at a flux density of 3.76~$\pm$~0.06 mJy. This corresponds to a spectral index of $\alpha=-$0.72~$\pm$0.02.  A few arcseconds away from E lies the western component located at R.A.~=~12$^{\rm h}$28$^{\rm m}$05.$^{\!\rm s}$02 ($\pm$0.$\!^{\rm s}$15), Decl.~=~$-$48\degr52\arcmin21\farcs15 ($\pm$0\farcs23) (J2000).  The spectral shape of W is steeper than that of E with a spectral index of $\alpha=-$1.09~$\pm$0.04 and with a flux density of 2.42~$\pm$~0.06 mJy at 1400 MHz, 0.55~$\pm$~0.03 mJy at 5.5 GHz and 0.31~$\pm$~0.03 mJy at 9 GHz.  

In observations 1 and 2 the ATCA antennae configurations were in a lower resolution mode and only detected a single component from J122806$-$485218.  During observation 1 the total source flux density was 2.4~$\pm$~0.2 mJy at 5.5 GHz and 1.2~$\pm$~0.1 mJy at 9 GHz, while during observation 2 the detected flux density was 3.7~$\pm$~0.6 mJy at 2.45 GHz.  Integrating the component flux densities in observation 3 yields a 5.5 GHz flux density of 2.39~$\pm$~0.05 mJy and 1.48~$\pm$~0.05 mJy at 9 GHz, which compares favourably with the measurements during observation 1 and gives no significant indication of radio variability.  In the GMRT observations at 640 MHz a single source is observed, as the beam size is as big as 24\arcsec~$\times$~27\arcsec. This is compatible with the location of the two components seen at other frequencies. The flux density at 640 MHz is 33.2~$\pm$~3.3 mJy, the integral flux density of the two components at 1400 MHz is 6.2~$\pm$~0.9 mJy, and no source is detected at 240 MHz.  

\begin{figure}
\centering
\includegraphics[width=0.95\linewidth,angle=0,clip=]{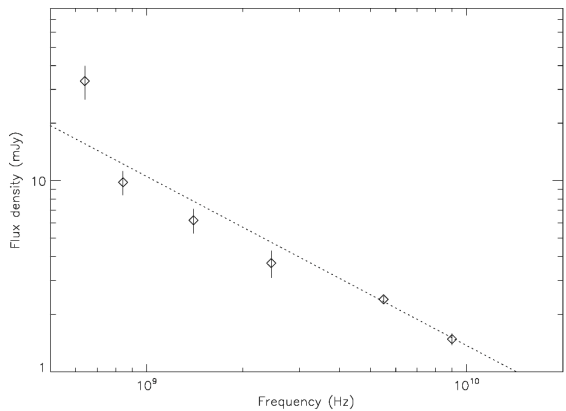}
\caption{The radio spectrum of the integrated components of the new radio source \atca.  The flux densities at 640 MHz and 1400 MHz from GMRT, at 843 MHz from MOST and at 2.45 GHz, 5.5 GHz and 9 GHz from ATCA are all plotted.  The vertical bars indicate the 1$\sigma$ errors on the data points and the dashed line indicates the linear best fit.}
\label{fig:radio-spec}
\end{figure}

The SUMMS survey catalogue is constructed from 843 MHz radio observations made by the Molongo Observatory Synthesis Telescope (MOST) and does not report a radio source at the location of J122806$-$485218.  A re-analysis of the MOST data indicates a radio source at the location of the western lobe. While it did not satisfy the independent criteria for inclusion into the SUMMS catalogue, in combination with the ATCA data we can confidently report a detection at 843 MHz.  The total flux density from the MOST data is 9.8~$\pm$~1.4 mJy at 843 MHz.  Combining the ATCA, GMRT and MOST data gives a spectral index of $\alpha=$-0.88~$\pm$~0.05 for the integrated flux density from all components in J122806$-$485218.  This is shown in Figure~\ref{fig:radio-spec}.  The spectral fit matches the radio flux density values well, with the exception of the 640 MHz point which is noticeably higher than would be expected; if the flux density is truly this strong at 640 MHz it implies that either we are resolving the source at 1400 MHz or a very steep component is present. 

\textbf{Source 3}: A third source is detected only at 5.5 GHz and only in observation 3. It is located at R.A.~=~12$^{\rm h}$27$^{\rm m}$35.$^{\!\rm s}$5, Decl.~=~$-$48\degr52\arcmin47\farcs2 (J2000) with an error of 0\farcs3.  The source is undetected in the 9 GHz observations and has a flux density of 0.23$\pm$0.02 mJy at 5.5 GHz.  Searching the \emph{XMM-Newton} observations of this field yields no significant detection of this source in X-rays.  

\begin{table}
\centering
\caption{Log details of the ATCA and GMRT radio observations of \ems.}
\begin{tabular}{|cccc|}
\hline
Obs. No. & Date & Time on source (h)  & Configuration \\
\hline
\hline
ATCA \\
\hline
1	&	02 Oct. 2009	&  2.37 & H75 @ 5.5 \& 9 GHz \\
2	&	13 Oct. 2009	&	 1.11 & H168 @ 2.45 GHz \\
3	&	06 Nov. 2009 & 7.10 &	1.5B @ 5.5 \& 9 GHz\\
\hline
GMRT \\
\hline
4 & 10 May 2010 & 2.07 & 240, 640 MHz \\
5 & 13 May 2010 & 2.18 & 1400 MHz\\
\hline
\end{tabular}
\label{tab:ATCA}
\end{table}

\subsection{Parkes radio observations}\label{parkes}
The $\gamma$-ray characteristics of \ems\ allow for the
possibility that it could be a rotation-powered pulsar.  In order to
investigate this, we searched the source for pulsed radio signals at
the Parkes telescope in Australia, first on November 25, 2009.

We used the central beam of the Parkes multibeam receiver centred
at a frequency of 1390\,MHz to record data for a total of 2.0\,hr.
The telescope was pointed at $\mbox{R.A.}=12^{\rm h}27^{\rm m}50.4^{\rm
s}$, $\mbox{Decl.}=-48^\circ51'54''$, with a beam $\mbox{FWHM}=14'$; the beam size is sufficient that the LAT source error ellipse is entirely covered at $>$90\% of the nominal sensitivity.
Total radio power was sampled every $125\,\mu$s from each of 512 channels
spanning a bandwidth of 256\,MHz \citep[more details of the observing setup
are given in, e.g.,][]{2001MNRAS.328...17M}.

The data were analyzed using standard pulsar search techniques implemented
in PRESTO \citep{RansomThesis}.  We searched a dispersion measure (DM) range of
0--270\,pc\,cm$^{-3}$, which comfortably includes the maximum possible
electron column density in this direction according to the \citet{2002astro.ph..7156C} model.  To allow for the possibility of a millisecond pulsar in a
tight binary, where the observed pulsar period $P$ varies due to Doppler
shifts, the search accounted for a large range of possible accelerations
(assumed constant, up to $\mbox{ZMAX=200}$ in PRESTO).  No significant pulsar candidates were identified.  

The nominal sensitivity limit of our search, assuming a pulse width of
$0.1\,P$, is approximately 0.06\,mJy for a long-period pulsar.  This limit
degrades rapidly for short (millisecond) periods, especially for high
values of $\mbox{DM}/P$.  Nevertheless, this search would have detected
most millisecond (or young) pulsars known in the disk of the Galaxy.
A potential caveat is that the received flux from a nearby pulsar
may vary greatly due to interstellar scintillation; binaries may also
experience eclipses.  We therefore repeated this observation twice, for
1.1\,hr on July 18, 2010, and 1.0\,hr on November 12, 2010.  We analyzed
the data in a similar manner, but again detected no pulsar signals.
There is therefore no evidence that 1FGL~J1227.9--4852 harbors a radio
pulsar beamed towards the Earth.

\subsection{ESO VLT \& archival optical \& IR observations}

A search of public archival observations finds no deep observations at ultraviolet, optical or infra-red wavelengths; there are no \emph{Spitzer} observations of this field. \xss\ has a 2MASS and DENIS counterpart with magnitudes ranging from 15.78 in the I band to 15.3 in the K$_{s}$ band; it should be noted that this source has been observed to flicker in the optical with variations of $>$1 magnitude on timescales of $<$15 minutes \citep{2009MNRAS.395..386P}.  There is no counterpart to the new radio source (\atca) in the 2MASS \citep{2006AJ....131.1163S} or DENIS catalogues at the position of the core up to K$_{s}>$14.3, H~$>$15.1, J~$>$15.8 and I~$>$18.5 magnitudes.

To identify the potential host galaxy of the new radio source \atca\ we performed imaging and spectroscopic observations of the field around R.A.~=~12$^{\rm h}$28$^{\rm m}$05\fs01, Dec.~=~$-$48\degr52\arcmin21\farcs3 (J2000) with the ESO Very Large Telescope FOcal Reducer and low dispersion Spectrograph (VLT FORS2).  Pre-imaging data in the R\_SPECIAL filter were collected on 2010 February 10; to detect candidate host galaxies we executed a set of four 15~s exposures.

Photometric observations were reduced for bias and flat-field using calibration frames available for the night of observations.
Magnitudes of isolated galaxies were measured using SExtractor \citep{1996A&AS..117..393B}. One of the galaxies in the field is blended with a neighboring star, to measure its magnitude we have fitted a PSF profile to the star and subtracted it from the image. The magnitudes of the galaxies were then measured using the elliptical aperture in SExtractor. From the galaxies in the field, two are the most probable candidates for the host galaxy:
\begin{itemize}
	\item Galaxy1 at R.A.~=~12$^{\rm h}$28$^{\rm m}$06\fs05, Dec.~=~$-$48\degr52\arcmin16\farcs32 (J2000) with R = 18.17$\pm$0.04 magnitudes is located less than 2\arcsec\ from the brighter eastern component of \atca.
	\item Galaxy2 at R.A.~=~12$^{\rm h}$28$^{\rm m}$05\fs42, Dec.~=~$-$48\degr52\arcmin18\farcs79 (J2000) with R = 20.19$\pm$0.12 magnitudes and lies between the eastern and western components.
\end{itemize}

In order to obtain the redshifts of the candidates, the following spectroscopic observations were conducted using the ESO VLT FORS2 in MXU mode and a mask with slits for 9 objects. We chose grism (300I) which samples wide wavelength range from 5000\AA\ to 11000\AA. A set of four 1200~s exposures was performed on 2010 July 15. Collected spectroscopic data were reduced twice using different tools; the ESO FORS pipeline and the NOAO IRAF package. Poor conditions and inadequate sky subtraction precluded sensitive measurements
of the of the two most probable host galaxies, mentioned above; it was not possible to extract 1D spectra with S/N ratios sufficient to identify any emission or absorption lines and to measure redshift.

\begin{figure}
\centering
\includegraphics[width=0.95\linewidth, clip]{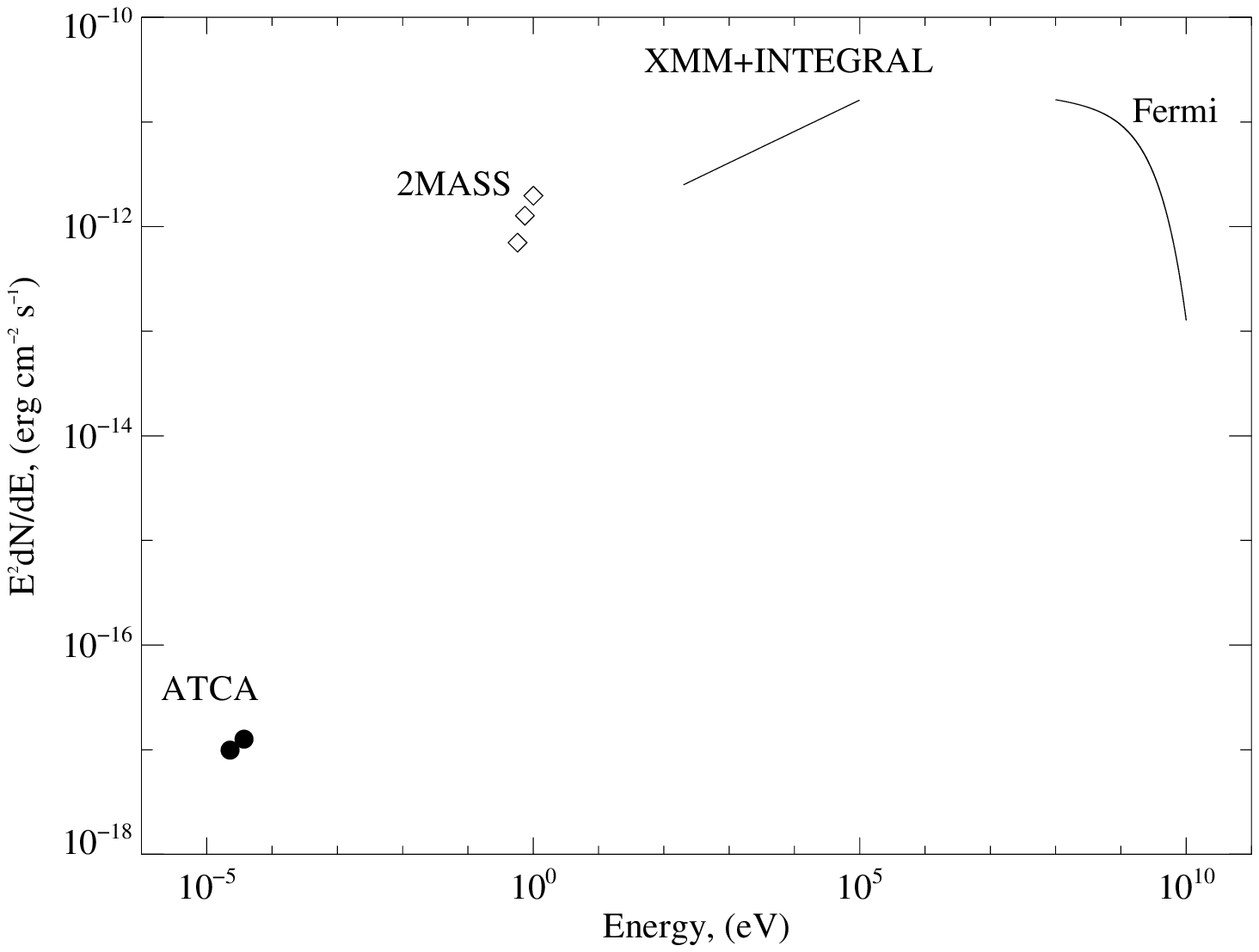}
\includegraphics[width=0.95\linewidth, clip]{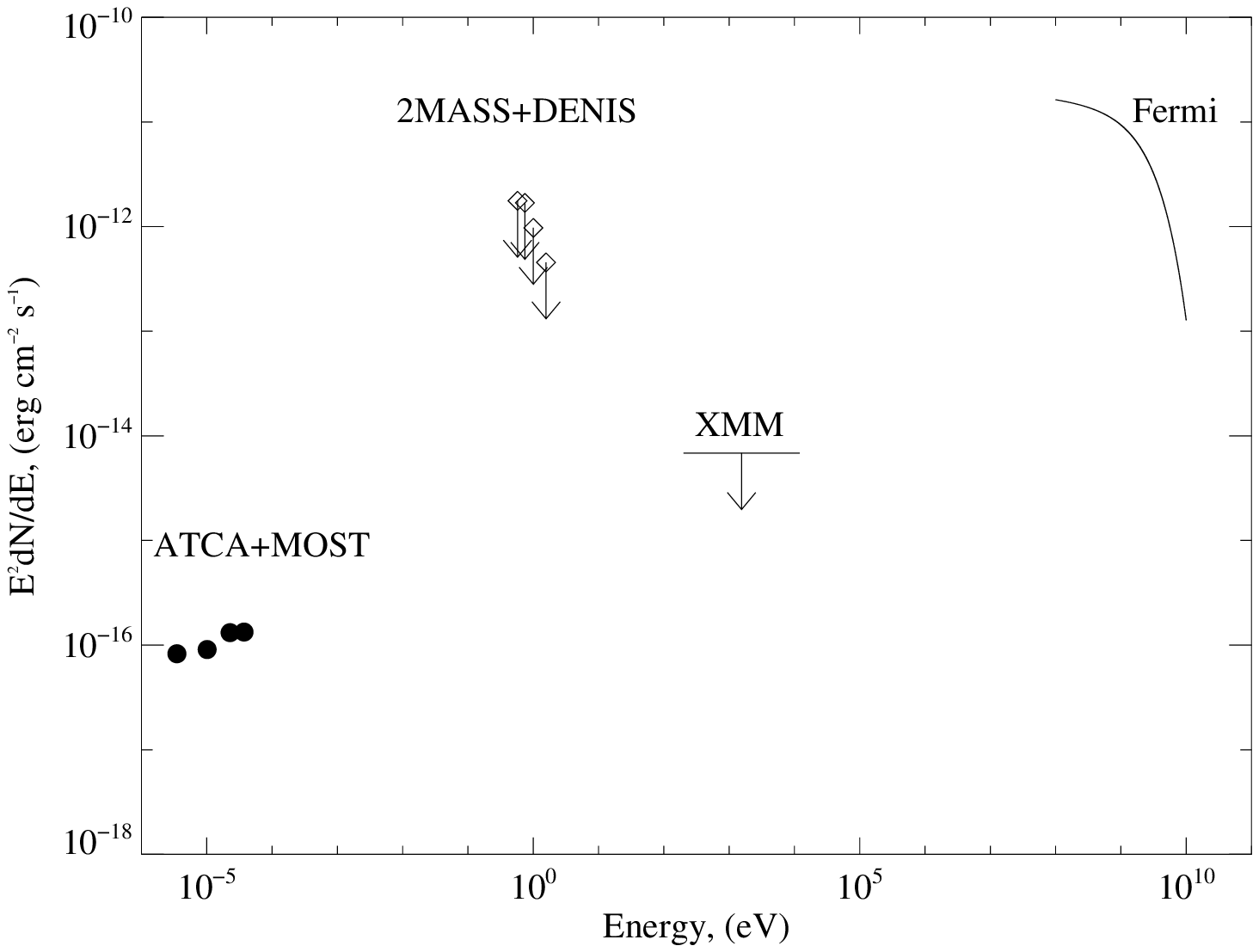}
\caption{\fermi\ LAT SED of 1FGL J1227.9$-$4852 overlaid on the overall SED of: (a) the low-mass X-ray binary \xss\ (top panel) and (b) the new compact double radio source discovered by ATCA, \atca\ (bottom panel). Radio, optical, X-ray and $\gamma$-ray bands are plotted.}
\label{fig:sed}
\end{figure}

\section{Discussion}

By assuming that \ems\ is associated with either the LMXB \xss\ or the new radio source \atca, we can combine the data from observations discussed in \S~2 to construct the broad-band spectral energy distribution shown in Figure~\ref{fig:sed}.  We calculated a 2$\sigma$ X-ray upper limit for \atca\ from the dataset analysed by \citet{2010A&A...515A..25D} by using the standard \emph{XMM-Newton} PPS sensitivity map  and assuming a simple power law with a photon index of 2. This gave a flux upper limit of $\sim$2$\times$10$^{-14}$erg~cm$^{-2}$~s$^{-1}$.

The observed characteristics of \ems\ do not allow a definitive source classification.  There is no clear indication of variability in the $\gamma$-ray band to be correlated with other wavelengths; and there is no significant periodic signal that could be attributed to a spin or orbit.  \ems\ is a persistent, stable emitter with an exponentially cutoff power law spectrum that lies approximately 14$^\circ$ off of the Galactic plane.  Within the LAT source error circle is a single bright, persistent X-ray source, the LMXB \xss, and a relatively bright, newly detected, persistent radio source \atca.  The possibility exists that \ems\ is related to neither of these multi-wavelength counterparts in which case little more can be said about its nature.  

To place this unidentified $\gamma$-ray source in context we must note that active galactic nuclei (AGN) constitute the largest group of identified \fermi\ sources ($\sim$80\%) \citep{1FGLcat, 1LAC} and the vast majority of these are seen as blazars.  Conversely, the Galactic source population is in the minority. It is principally composed of rotation-powered pulsars with small numbers of high-mass X-ray binaries, supernova remnants, globular clusters and pulsar wind nebulae. The pulsars all exhibit exponentially cutoff power law spectra and are identified through the detection of pulsations in the $\gamma$-ray.

\subsection{Is the $\gamma$-ray source Galactic in nature?}
\subsubsection{An isolated pulsar?}\label{isoPulsar}
The $\gamma$-ray characteristics of \ems\ are highly reminiscent of the population of $\gamma$-ray pulsars which have been identified in the first few years of observations of \fermi-LAT \citep{HEEPPulsars}.  While the photon index $\Gamma$=2.20 and cutoff energy of 4 GeV are rather steep and high respectively compared to the typical LAT detected pulsars \citep{HEEPPulsars} there are two LAT detected $\gamma$-ray pulsars with photon indices $>$2.2 \citep[PSR J0659+1414 and PSR J1833$-$1034, see][]{2010ApJS..187..460A}.  Additionally there are LAT-discovered millisecond pulsars (MSPs) that share similar characteristics; PSR J0218+4232 and PSR J0437$-$4715 have photon indices of 2.0 and 2.1 and cutoff energies of 7 GeV and 2.1 GeV respectively \citep{2009Sci...325..848A}.  The location of \ems\ off the Galactic plane would suggest that if it is a pulsar then it would likely belong to the population of MSPs.  However, searches for pulsations in both $\gamma$-rays with the LAT and in radio with Parkes have not yielded any significant detection (see \S~\ref{timing}). The Parkes searches were sensitive to almost any known radio millisecond pulsar beamed towards the Earth, and there is therefore no direct evidence that \ems\ contains such a pulsar.  The low flux of 0.18 mJy at 5.5 GHz of \xss\ is consistent with what would be expected from an isolated pulsar.  The spectral index is $0.5~\pm~0.6$, which is not as steep as would be expected from a pulsar \citep[see e.g.,][]{1999ApJ...526..957K, 1998A&AS..127..153K}.  While this does not rule out an isolated pulsar nature of the source, it does decrease its probability.

\subsubsection{A pulsar in a binary?}
Binary evolution theory explains the origins of MSPs as `recycled pulsars', pulsars that once had a much longer spin period and had a low-mass binary companion.  The pulsar accretes matter through Roche-lobe overflow and is seen as a bright X-ray source, an LMXB.  Through the accretion process angular momentum is transferred to the pulsar, effectively spinning it up to millisecond periods.  Once the accretion rate drops low enough magnetic pressure can be sufficient to prevent further matter being accreted, at which point the rotation-powered radio pulsations can be seen.

Targeted radio follow-ups of $>$100 of the unidentified 1FGL sources have discovered 18 new MSPs in addition to the 11 radio MSPs detected by the LAT \citep{HEEPPulsars}. If the $\gamma$-ray source was associated with a pulsar in a binary, possibly the LMXB \xss, then the modulation of the binary orbit would make searching for pulsations much more difficult due to the combination of the orbital motion and potential eclipses.  

Interpreting \xss\ as a millisecond binary pulsar explains the observed short orbital period, low X-ray luminosity with eclipses and flares, optical modulation from the irradiated companion and the presence of radio emission \citep[see e.g. ][for PSR J0024$-$7204W in 47 Tuc]{2005ApJ...630.1029B}.  Since GeV emission has been detected from millisecond binary pulsars then this interpretation for \xss\ also provides a coherent picture of the multi-wavelength observations and could associate it with the GeV source \ems.  It should be noted that of the binary pulsars that are LAT sources, none show GeV modulation at their orbital periods. This hypothesis presents two puzzles:
\begin{enumerate}[1.]
	\item The radio spectrum is not as steep as would be expected but this could be a result of free-free absorption making the spectrum flatter.
	\item The optical spectroscopy shows emission lines \citep{2006A&A...459...21M, 2009MNRAS.395..386P} which are typical of accretion. However, all the optical spectroscopic measurements pre-date the launch of \textit{Fermi} and since then the system may have transitioned from the accreting LMXB state to the rotation-powered pulsar phase. 
\end{enumerate}

PSR J1023+0038 is a system in which \citet{2009Sci...324.1411A} suggest this has recently happened.  \citet{2009Sci...324.1411A}; \citet{2009ApJ...703.2017W} discovered millisecond radio pulsations in the system, however historical optical observations were very different from contemporary observations.  The historic optical observations were very blue and exhibited rapid flickering of $\sim$1 magnitude indicative of an accretion flow; this behaviour is similar to what has been seen in \xss\ \citep{2010A&A...515A..25D}.  Recently \citet{2010arXiv1010.4311T} reported the detection of high energy $\gamma$-ray emission at the 7$\sigma$ level at a position consistent with PSR J1023+0038; no pulsations were detected in the $\gamma$-ray data but a steep $\gamma$-ray spectrum with a power law photon index of $\sim$2.9 was observed.  Investigating the $\gamma$-ray to X-ray luminosity ratio in PSR J1023+0038 and \xss\ we find that both sources are a factor $\sim$10 brighter in $\gamma$-rays.

Could \xss\ be a system similar to PSR J1023+0038?  If \ems\ is indeed a MSP with a short orbital period, this could explain why pulsations have yet to be detected from the source.  The absence of any recent optical spectroscopy of \xss\ means that we do not know if it is still an accreting source in the \textit{Fermi} era. If new optical spectra were to show no signatures of accretion then the case for \xss\ being a millisecond binary pulsar would be significantly strengthened.  The hard X-rays detected by \emph{INTEGRAL} would be unexpected in the case of a non-accreting MSP. However \xss\ is reported as a weak persistent source in the latest catalogue \citep{2010ApJS..186....1B} with a detection level of 10.2$\sigma$ in $\sim$1 Ms of exposure and it should be noted that all of the data used precedes the launch of \fermi\ and could have been while the source was acting as an accreting LMXB. An interesting historical note is that when PSR J1023+0038 was initially discovered it was believed to be the first radio selected cataclysmic variable \citep{2005AJ....130..759T, 2002PASP..114.1359B} in much the same way that \xss\ was initially classified as a CV.

\subsection{Is the $\gamma$-ray source extragalactic in nature?}
The largest population of associated high-latitude ($b>10^{\circ}$) LAT sources are AGN, all of which are radio sources at some level \citep{1LAC}.  Hence, despite \atca\ being at the edge of the LAT source 99.7\% error circle, as it is by far the brightest potential radio counterpart and has an AGN-like morphology it initially appears to be a strong candidate as the counterpart to \ems.  The new radio source, \atca, has a double morphology often seen in radio-loud AGN; it can be identified as either a core/jet structure (with the brightest component marking the position of the radio core and the weaker one being a hotspot in a jet), or as a roughly symmetric pair of radio lobes separated by $\sim$10\arcsec.  The core has a spectral index of $\alpha$=$-$0.72, while the lobe has a much steeper index of $\alpha$=$-$1.09.  Hence if \ems\ is associated with this object then it appears to be of an extragalactic nature.

The majority of the AGN detected by the LAT are blazars. These are AGN where the line of sight is aligned with the jet axis. They are radio bright, variable and with flat or inverted ($\alpha$ $\geq$ -0.5) spectra. Radio galaxies are observed at higher angles and are generally less luminous. Their radio spectra are also steeper ($\alpha$ $\leq$ -0.5).  The absence of $\gamma$-ray variability and a steep radio spectrum would suggest that \atca\ does not belong to the blazar class.

The radio galaxy Centaurus A has also been seen as an extragalactic source of $\gamma$-ray emission. \citet{2010Sci...328..725F} report the detection of HE $\gamma$-rays specifically from the giant radio lobes of Cen A and that the lobe flux contributed the majority of the emission.  This was interpreted as inverse Compton scattered radiation from the cosmic microwave background (CMB), with additional contribution at higher energies from the infrared-to-optical extragalactic background light (EBL).  Could the radio lobe in \atca\ also upscatter CMB photons to $\gamma$-ray energies creating the emission of \ems? The lack of $\gamma$-ray variability would be expected in such a case and this scenario can give a large $\gamma$ to radio ratio at high redshift because lobe electrons upscatter CMB photons whose energy density increases as \mbox{(1 + z)$^4$}.  The ratio of the $\gamma$ to radio luminosities is proportional to the ratio of energy densities in the magnetic (B) field and CMB and is shown below:

\begin{eqnarray*}
\frac{L_{Radio}}{L_{\gamma}} \propto \frac{U_B}{U_{CMB}} & = & \frac{B^2/8\pi}{4\times10^{-13}\ (1+z)^4} \\
	& = & 10^{-5}\\
\Rightarrow B & \sim & 10^{-8}\ (1+z)^2  \ \ {\rm Gauss}
\label{LumRatio}
\end{eqnarray*}
 
This is much lower than the $\mu$Gauss magnetic fields typically invoked for extended radio structures in jetted AGN \citep[e.g.][]{2005ApJ...622..797K}. One explanation is that the synchrotron emission is vastly underestimated and would require there to be large scale diffuse emission. However there is no evidence of large scale diffuse emission in the GMRT observations, effectively ruling out the upscattering of CMB photons in the radio lobe of \atca\ as a viable explanation for the $\gamma$-ray emission of \ems.  This conclusion is further supported by the particularly high $\gamma$-ray to X-ray ratio implied by the XMM upper limit as copious X-rays would be generated via the same processes as $\gamma$-rays within the lobes.

\section{Conclusions}
The unidentified \fermi\ source \ems\ is an intriguing source that may potentially be associated with either the bright X-ray source, \xss\ or the new radio source \atca, which have been identified within the LAT error circle.  However, the specific nature of this object is still unclear and it may not be firmly associated with either of these two objects yet.  The emission is unlikely to be from a blazar jet as no flat-spectrum radio source is identified.  Investigating the possibility that the radio lobe of \atca\ can upscatter CMB photons to high energies implies that this is a very unlikely scenario based upon the observed radio and $\gamma$-ray luminosities.  The source may be yet another example of the radio-faint/quiet $\gamma$-ray pulsars detected by \fermi, with the non-detection indicating that current $\gamma$-ray pulsation searches have not been sensitive enough.  The blind searches carried out to date on the LAT data would not have detected either an MSP or a binary pulsar.

Based upon the available observations, we favour the hypothesis that \ems\ is associated with \xss\ as the least improbable association, in the scenario that the system has evolved from an accreting LMXB phase to a millisecond binary pulsar state. If this is the case it would be an exciting link in understanding binary evolution.  New optical spectroscopic measurements of \xss\ will indicate whether it is still an accreting source or whether it may have transitioned to be a new millisecond binary pulsar.  Further searches for a new pulsar in the field will allow for a more confident assessment of the likelihood of pulsar origin of the $\gamma$-rays.  Another key to identifying the nature of this source may be in correlated variability at multiple wavelengths.  However, to date there has been no indication of any variability in the GeV band.  Of course, the possibility remains that \ems\ is something entirely different, as the source has not as yet been associated with a multi-wavelength counterpart.    

\section*{Acknowledgments}
The \textit{Fermi} LAT Collaboration acknowledges generous ongoing support
from a number of agencies and institutes that have supported both the
development and the operation of the LAT as well as scientific data analysis.
These include the National Aeronautics and Space Administration and the
Department of Energy in the United States, the Commissariat \`a l'Energie Atomique
and the Centre National de la Recherche Scientifique / Institut National de Physique
Nucl\'eaire et de Physique des Particules in France, the Agenzia Spaziale Italiana
and the Istituto Nazionale di Fisica Nucleare in Italy, the Ministry of Education,
Culture, Sports, Science and Technology (MEXT), High Energy Accelerator Research
Organization (KEK) and Japan Aerospace Exploration Agency (JAXA) in Japan, and
the K.~A.~Wallenberg Foundation, the Swedish Research Council and the
Swedish National Space Board in Sweden.

Additional support for science analysis during the operations phase is gratefully
acknowledged from the Istituto Nazionale di Astrofisica in Italy and the Centre National d'\'Etudes Spatiales in France.

We thank the staff of the GMRT who have made these observations possible. GMRT is run by the National Centre for Radio Astrophysics of the Tata Institute of Fundamental Research.  The Australia Telescope is funded by the Commonwealth of Australia for operation as a national Facility managed by CSIRO.  Based on observations made with ESO Telescopes at the La Silla Observatory under programmes ID 284.B-5030 and 285.B-5020.  ABH, GD and AS acknowledge funding by contract ERC-StG-200911 from the European Community. The research leading to these results has received partial funding from the European Community's Seventh Framework Programme (FP7/2007-2013) under grant agreement number ITN 215212 ``Black Hole Universe''.


\begin{thebibliography}{99}

\bibitem[\protect\citeauthoryear{Abdo et al.}{2010a}]{1FGLcat} 
Abdo A.~A., et al., 2010a, ApJS, 188, 405 


\bibitem[\protect\citeauthoryear{Abdo et al.}{2010b}]{1LAC} 
Abdo A.~A., et al., 2010b, ApJ, 715, 429 

\bibitem[\protect\citeauthoryear{Abdo et al.}{2010c}]{2010Sci...328..725F} 
Abdo A.~A., et al., 2010c, Sci, 328, 725 

\bibitem[\protect\citeauthoryear{Abdo et al.}{2010d}]{2010ApJS..187..460A} 
Abdo A.~A., et al., 2010d, ApJS, 187, 460 

\bibitem[\protect\citeauthoryear{Abdo et al.}{2009a}]{2009Sci...326.1512F} 
Abdo A.~A., et al., 2009a, Sci, 326, 1512 

\bibitem[\protect\citeauthoryear{Abdo et al.}{2009b}]{2009ApJ...706L..56A} 
Abdo A.~A., et al., 2009b, ApJ, 706, L56 

\bibitem[\protect\citeauthoryear{Abdo et al.}{2009c}]{2009Sci...325..840A} 
Abdo A.~A., et al., 2009c, Sci, 325, 840 

\bibitem[\protect\citeauthoryear{Abdo et al.}{2009d}]{2009Sci...325..848A} 
Abdo A.~A., et al., 2009d, Sci, 325, 848 

\bibitem[\protect\citeauthoryear{Abdo et al.}{2009e}]{2009ApJ...701L.123A} 
Abdo A.~A., et al., 2009e, ApJ, 701, L123 

\bibitem[\protect\citeauthoryear{Abdo et al.}{2009f}]{2009ApJ...699..817A} 
Abdo A.~A., et al., 2009f, ApJ, 699, 817 

\bibitem[\protect\citeauthoryear{Archibald et al.}{2009}]{2009Sci...324.1411A} 
Archibald A.~M., et al., 2009, Sci, 324, 1411 

\bibitem[\protect\citeauthoryear{Atwood et al.}{2009}]{2009ApJ...697.1071A} 
Atwood W.~B., et al., 2009, ApJ, 697, 1071 

\bibitem[\protect\citeauthoryear{Atwood et al.}{2006}]{2006ApJ...652L..49A} 
Atwood W.~B., Ziegler M., Johnson R.~P., Baughman B.~M., 2006, ApJ, 652, 
L49 

\bibitem[\protect\citeauthoryear{Bertin \& Arnouts}{1996}]{1996A&AS..117..393B} 
Bertin E., Arnouts S., 1996, A\&AS, 117, 393

\bibitem[\protect\citeauthoryear{Bird et al.}{2010}]{2010ApJS..186....1B} 
Bird A.~J., et al., 2010, ApJS, 186, 1 

\bibitem[\protect\citeauthoryear{Bogdanov, Grindlay, 
\& van den Berg}{2005}]{2005ApJ...630.1029B} Bogdanov S., Grindlay J.~E., van den Berg M., 2005, ApJ, 630, 1029 

\bibitem[\protect\citeauthoryear{Bond et al.}{2002}]{2002PASP..114.1359B} 
Bond H.~E., White R.~L., Becker R.~H., O'Brien M.~S., 2002, PASP, 114, 1359 


\bibitem[\protect\citeauthoryear{Butters et 
al.}{2008}]{2008A&A...487..271B} Butters O.~W., Norton A.~J., Hakala P., Mukai K., Barlow E.~J., 2008, A\&A, 487, 271 

\bibitem[\protect\citeauthoryear{Cordes 
\& Lazio}{2002}]{2002astro.ph..7156C} Cordes J.~M., Lazio T.~J.~W., 2002, arXiv:astro-ph/0207156 

\bibitem[\protect\citeauthoryear{Cusumano et 
al.}{2010}]{2010A&A...510A..48C} Cusumano G., et al., 2010, A\&A, 510, A48 

\bibitem[\protect\citeauthoryear{de Martino et 
al.}{2010}]{2010A&A...515A..25D} de Martino D., et al., 2010, A\&A, 515, A25 

\bibitem[\protect\citeauthoryear{Kataoka 
\& Stawarz}{2005}]{2005ApJ...622..797K} Kataoka J., Stawarz {\L}., 2005, ApJ, 622, 797

\bibitem[\protect\citeauthoryear{Kijak et al.}{1998}]{1998A&AS..127..153K} 
Kijak J., Kramer M., Wielebinski R., Jessner A., 1998, A\&AS, 127, 153 


\bibitem[\protect\citeauthoryear{Kramer et al.}{1999}]{1999ApJ...526..957K} 
Kramer M., Lange C., Lorimer D.~R., Backer D.~C., Xilouris K.~M., Jessner A., Wielebinski R., 1999, ApJ, 526, 957 

\bibitem[{{Lomb}(1976)}]{1976Ap&SS..39..447L}
{Lomb}, N.~R. 1976, Ap\&SS, 39, 447


\bibitem[\protect\citeauthoryear{Manchester et 
al.}{2001}]{2001MNRAS.328...17M} Manchester R.~N., et al., 2001, MNRAS, 
328, 17 

\bibitem[\protect\citeauthoryear{Masetti et 
al.}{2006}]{2006A&A...459...21M} Masetti N., et al., 2006, A\&A, 459, 21 

\bibitem[\protect\citeauthoryear{Mattox et al.}{1996}]{1996ApJ...461..396M} 
Mattox J.~R., et al., 1996, ApJ, 461, 396 

\bibitem[\protect\citeauthoryear{Mauch et al.}{2003}]{2003MNRAS.342.1117M} 
Mauch T., Murphy T., Buttery H.~J., Curran J., Hunstead R.~W., Piestrzynski 
B., Robertson J.~G., Sadler E.~M., 2003, MNRAS, 342, 1117 

\bibitem[\protect\citeauthoryear{Pretorius}{2009}]{2009MNRAS.395..386P} 
Pretorius M.~L., 2009, MNRAS, 395, 386 


\bibitem[\protect\citeauthoryear{Ransom}{2001}]{RansomThesis}
{Ransom} S., 2001, PhD thesis, Harvard University

\bibitem[\protect\citeauthoryear{Ray \& Saz Parkinson}{2010}]{HEEPPulsars}
{Ray} P. \& Saz Parkinson P.~M., 2010, in Proc. - ``ICREA Workshop on The High-Energy Emission from Pulsars and their Systems'', Barcelona, in press, arXiv:1007.2183

\bibitem[\protect\citeauthoryear{Saitou et al.}{2009}]{2009PASJ...61L..13S} 
Saitou K., Tsujimoto M., Ebisawa K., Ishida M., 2009, PASJ, 61, L13 

\bibitem[\protect\citeauthoryear{Sault \& Killeen}{2010}]{MiriadUG} 
Sault R.~J. \& Killeen N.~E.~B., 2010, The Miriad User's Guide, Sydney: Australia Telescope National Facility

\bibitem[\protect\citeauthoryear{Sault, Teuben, \& Wright}{1995}]{1995ASPC...77..433S} 
Sault R.~J., Teuben P.~J., Wright M.~C.~H., 1995, ASPC, 77, 433 

\bibitem[\protect\citeauthoryear{Sault \& Wieringa}{1994}]{1994A&AS..108..585S} 
Sault R.~J., Wieringa M.~H., 1994, A\&AS, 108, 585 

\bibitem[\protect\citeauthoryear{Saz Parkinson et al.}{2010}]{2010ApJ...725..571S} 
Saz Parkinson, P.~M., et al., 2010, ApJ, 725, 571

\bibitem[\protect\citeauthoryear{Scargle}{1982}]{1982ApJ...263..835S}
{Scargle}, J.~D. 1982, ApJ, 263, 835

\bibitem[\protect\citeauthoryear{Schoenmakers et 
al.}{2000}]{2000MNRAS.315..395S} Schoenmakers A.~P., de Bruyn A.~G., 
R{\"o}ttgering H.~J.~A., van der Laan H., 2000, MNRAS, 315, 395 

\bibitem[\protect\citeauthoryear{Skrutskie et 
al.}{2006}]{2006AJ....131.1163S} Skrutskie M.~F., et al., 2006, AJ, 131, 
1163 

\bibitem[\protect\citeauthoryear{Tam et al.}{2010}]{2010arXiv1010.4311T} 
Tam P.~H.~T., et al., 2010, ApJ, 724, L207 

\bibitem[\protect\citeauthoryear{Thorstensen 
\& Armstrong}{2005}]{2005AJ....130..759T} Thorstensen J.~R., Armstrong E., 2005, AJ, 130, 759 

\bibitem[\protect\citeauthoryear{Ubertini et 
al.}{2009}]{2009ApJ...706L...7U} Ubertini P., Sguera V., Stephen J.~B., 
Bassani L., Bazzano A., Bird A.~J., 2009, ApJ, 706, L7 

\bibitem[\protect\citeauthoryear{Wang et al.}{2009}]{2009ApJ...703.2017W} 
Wang Z., Archibald A.~M., Thorstensen J.~R., Kaspi V.~M., Lorimer D.~R., 
Stairs I., Ransom S.~M., 2009, ApJ, 703, 2017 


\end{thebibliography}

\end{document}